# The computational simulation of the reflection spectra of copper red glaze

Gen Li, Yong Lei[*]

*Department of Conservation Science and Technology, The Palace Museum, Beijing 100009, People's Republic of China*

Correspondence to:

Yong Lei, Palace Museum, Beijing 100009, People's Republic of China. Tel: +86 010-85007260; e-mail: leiyong@dpm.org.cn

**ABSTRACT**

Owing to the limitation of traditional analytical methods, the coloration mechanism of copper red glaze has been disputed in the academic field for a long time, which mainly focuses on whether the color agent is metallic copper nanoparticles or cuprous oxide ($Cu_2O$) nanoparticles. Based on Mie scattering theory, this work calculated the reflection spectra of nanoparticles uniformly dispersed in transparent glaze with different types, diameters and volume fractions, then discussed the differences between the reflection spectra of metallic copper and cuprous oxide as scatterers, calculated the corresponding L*a*b* values, and compared them with the experimental results. This work provides a feasible and convenient method to distinguish these two coloration mechanisms.

**KEY WORDS:** copper red glaze; reflection spectra; Mie scattering

## I. INTRODUCTION

As a famous kind of porcelain in ancient China, copper red glaze has always been famous for its beautiful appearance and high firing difficulty, and also received much attention from the academic community. Generally speaking, the copper red glaze was first produced in Changsha Kiln in Tang Dynasty, but at that time the red glaze was mainly used as local decoration because the kilnmen did not have the technique to control the formation of copper red glaze perfectly. After the development of several hundred years, genuine copper-red glaze was thought to emerge in the late Yuan Dynasty, and reach its peak in the early Ming Dynasty, especially during the reign of Yongle and Xuande, and some famous types of products were invented, such as bright red and sacrificial red glaze. When it comes to the Qing Dynasty, the techniques of firing perfect copper red glaze were lost, so the kilnmen invented new recipes to imitate the artefacts in Ming Dynasty and produce several new types of copper red glaze, such as peach bloom glaze and Lang Kiln red glaze. Fig. 1 shows two typical copper red glaze artefacts of ancient China, namely the monk hat pot with bright red glaze produced in Ming Dynasty, and the plum vase with sacrificial red glaze produced in Qing Dynasty.

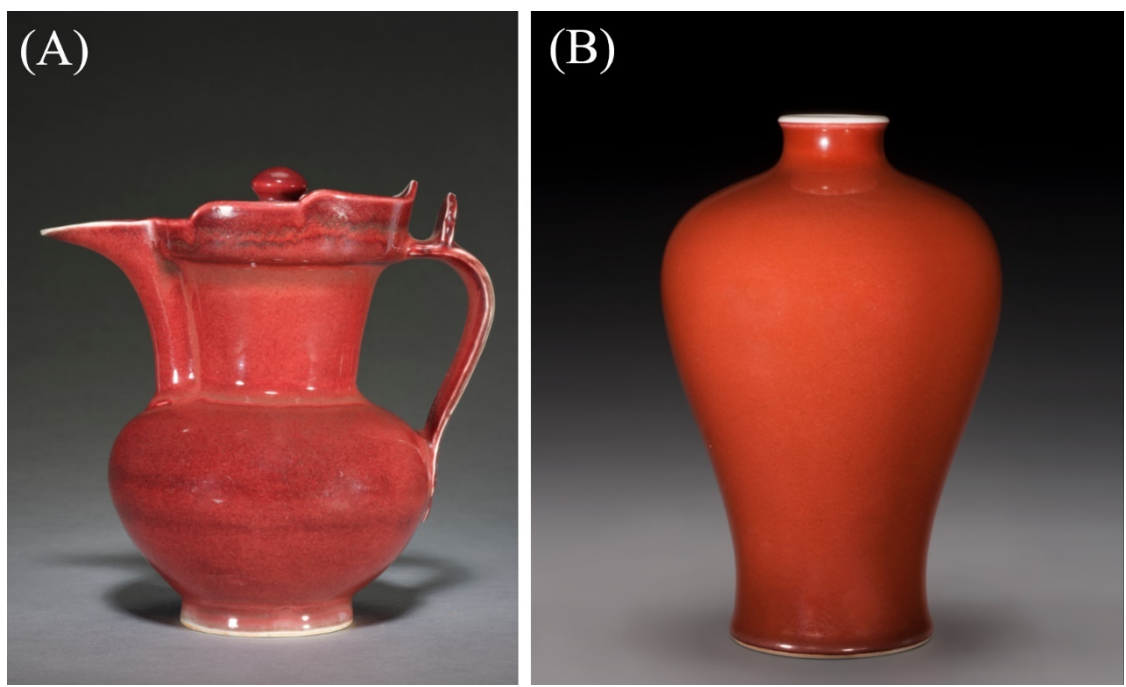

Fig.1 (a) Monk hat pot with bright red glaze produced during the Xuande reign period, Ming Dynasty; (b) Plum vase with sacrificial red glaze produced during the Yongzheng reign period, Qing Dynasty. (images come from https://digicol.dpm.org.cn/?category=6&color=ff3333)

The current research of copper red glaze mainly focuses on its firing process, raw material source, coloration mechanism, etc. In the study of coloration mechanism, the consensus of the academic community is that the red color comes from copper-containing nanoparticles in the glaze but not copper ions, because $Cu^+$ ion is almost colorless, and $Cu^{2+}$ ion generally appears green. On the other hand, these particles may be metallic copper or cuprous oxide ($Cu_2O$), because both of them are red in the bulk state, but which one dominates is still under debate. Nowadays more scholars tend to think that the colorant in copper red glaze is mainly metallic copper nanoparticles[1-3], but scholars who think cuprous oxide nanoparticles are the main colorant also give some evidence[4,5]. In addition, for ancient ceramics with complex hues such as Jun kiln porcelain, some scholars also analyze it in combination with the microstructure inside the glaze layer, such as liquid-liquid phase separation to form small droplets of the second phase, and regard its hue as the joint action of nanostructures and metal ions[6].

Generally speaking, the analysis methods of copper red glaze can be summarized as follows: XRD is used to analyze the phases[1,7], XRF and EDS are used to analyze the type and content of elements[1,7,8], XPS[9,10] and XANES[11-15] are used to analyze the valence state, and then at the micro level, SEM and TEM are used to observe the sizes and phases of particles[1,3,6], while EXAFS is used to study the coordination relationship of atoms[2,16]. These methods have greatly promoted peoples understanding of copper red glaze, but they are still insufficient to solve the problem of coloration mechanism.

For example, the spatial resolution of conventional chemical analytical methods such as XPS and XRF are confined to micrometers, which makes the signal collected come from not only nanoparticles but also the transparent glaze. In fact, according to previous reference, $Cu^+$ ion also exists in transparent glaze, and for typical sacrificial red glaze, the content of copper in the colorless layer is even higher than that in the red layer[12]. From this point of view, even if XPS or XRF gives the signal of $Cu^0$ in the red layer, we cannot assert how many percent of nanoparticles are metallic copper nanoparticles. Electron microscopy can accurately determine the size and phase structure of nanoparticles, but the number of observed particles is small and the statistics is not strong enough. In addition, these methods above are all invasive because the cross-section sample is needed in the study of copper red glaze.

Reflection spectrum is indeed a genuine non-invasive method. However, if we want to use reflection spectrum as an indicator of metallic copper or cuprous oxide nanoparticles, two standard samples with pure metallic copper/cuprous oxide nanoparticles uniformly dispersed are necessary to use their reflection spectra as reference, but producing such a glaze sample with homogeneously dispersed $Cu_2O$ nanoparticles is difficult. To sum up, it is not so easy to quickly determine whether a given copper red glaze sample, especially an unbroken artifact is colored by metallic copper particles or cuprous oxide particles.

When the experimental methods meet the bottleneck, the theoretical computational simulation often gives us new inspiration. Previously, Pradell et al. calculated the absorption and scattering cross-sections of nanoparticles based on Mie scattering theory, and based on that discussed the optical properties of the system[17]; Cuvelier et al. used the four-flux model to discuss the influence of copper particles with different radius and different volume fraction on coloration, and also calculated the scattering behavior of the system when cuprous oxide was coated on the surface of metallic copper particles[18]. This paper will start from Mie scattering theory to establish a numerical calculation framework based on multiple scattering, and simulate the reflection spectrum of transparent glaze with uniformly dispersed nanoparticles of different sizes, different volume fractions and different materials. Then the authors will discuss the difference of reflection spectrum between systems with metallic copper and cuprous oxide particles as colorants, and provide a new perspective for the coloration of copper red glaze.

**II. COMPUTATIONAL METHOD**

2.1 Basic theory of Mie scattering

The Mie scattering theory of spherical scatterers will be introduced in this section, which is mainly summarized from Bohren & Huffman[19] and the manuscript of Mätzler[20].

2.1.1 Expansion of incident and scattered wave

Consider the incoming monochromatic plane wave on the surface of a sphere:

$$\mathbf{E}_i = E_0 e^{i\mathbf{k}\cdot\mathbf{r}\cos\theta}\hat{\mathbf{e}}_x,$$

In the inner and outside part of this sphere, the electromagnetic field should satisfy the Maxwell equations. Using spherical coordinate system, the incoming plane wave can be written as the summation of a series of spherical harmonic function:

$$\mathbf{E}_i = E_0 \sum_{n=1}^{\infty} i^n \frac{2n+1}{n(n+1)} \left( \mathbf{M}_{o1n}^{(1)} - i\mathbf{N}_{e1n}^{(1)} \right)$$

The inner and scattered electromagnetic field of the sphere can be written in the following form:

$$\mathbf{E}_l = \sum_{n=1}^{\infty} E_n \left( c_n \mathbf{M}_{o1n}^{(1)} - id_n \mathbf{N}_{e1n}^{(1)} \right), \quad \mathbf{H}_l = \frac{-k_l}{\omega \mu_l} \sum_{n=1}^{\infty} E_n \left( d_n \mathbf{M}_{e1n}^{(1)} + ic_n \mathbf{N}_{o1n}^{(1)} \right),$$

$$\mathbf{E}_s = \sum_{n=1}^{\infty} E_n \left( ia_n \mathbf{N}_{e1n}^{(3)} - b_n \mathbf{M}_{o1n}^{(3)} \right), \quad \mathbf{H}_s = \frac{k}{\omega \mu} \sum_{n=1}^{\infty} E_n \left( ib_n \mathbf{N}_{o1n}^{(3)} + a_n \mathbf{M}_{e1n}^{(3)} \right).$$

where

$$\mathbf{M}_{o1n} = \cos\phi\, \pi_n(\cos\theta)\, z_n(\rho)\, \hat{\mathbf{e}}_\theta - \sin\phi\, \tau_n(\cos\theta)\, z_n(\rho)\, \hat{\mathbf{e}}_\phi,$$

$$\mathbf{M}_{e1n} = -\sin\phi\, \pi_n(\cos\theta)\, z_n(\rho)\, \hat{\mathbf{e}}_\theta - \cos\phi\, \tau_n(\cos\theta)\, z_n(\rho)\, \hat{\mathbf{e}}_\phi,$$

$$\mathbf{N}_{o1n} = \sin\phi\, n(n+1) \sin\theta\, \pi_n(\cos\theta) \frac{z_n(\rho)}{\rho} \hat{\mathbf{e}}_r$$
$$+ \sin\phi\, \tau_n(\cos\theta) \frac{[\rho z_n(\rho)]'}{\rho} \hat{\mathbf{e}}_\theta + \cos\phi\, \pi_n(\cos\theta) \frac{[\rho z_n(\rho)]'}{\rho} \hat{\mathbf{e}}_\phi,$$

$$\mathbf{N}_{e1n} = \cos\phi\, n(n+1) \sin\theta\, \pi_n(\cos\theta) \frac{z_n(\rho)}{\rho} \hat{\mathbf{e}}_r$$
$$+ \cos\phi\, \tau_n(\cos\theta) \frac{[\rho z_n(\rho)]'}{\rho} \hat{\mathbf{e}}_\theta - \sin\phi\, \pi_n(\cos\theta) \frac{[\rho z_n(\rho)]'}{\rho} \hat{\mathbf{e}}_\phi.$$

and we have

$$\pi_n = \frac{P_n^1}{\sin\theta}, \quad \tau_n = \frac{dP_n^1}{d\theta}.$$

For the inner magnetic field, $z_n$ stands for spherical Bessel function $j_n(k_1 r)$; and for the scattered magnetic field, it stands for spherical Bessel function $h_n^{(1)}(kr)$:

$$j_n(\rho) = \sqrt{\frac{\pi}{2\rho}} J_{n+1/2}(\rho), \quad y_n(\rho) = \sqrt{\frac{\pi}{2\rho}} Y_{n+1/2}(\rho),$$

$$h_n^{(1)}(\rho) = j_n(\rho) + iy_n(\rho), \quad h_n^{(2)}(\rho) = j_n(\rho) - iy_n(\rho).$$

and the four series of scattering parameters can be written as:

$$a_n = \frac{m^2 j_n(mx)[xj_n(x)]' - j_n(x)[mxj_n(mx)]'}{m^2 j_n(mx)[xh_n^{(1)}(x)]' - h_n^{(1)}(x)[mxj_n(mx)]'};$$

$$b_n = \frac{j_n(mx)[xj_n(x)]' - j_n(x)[mxj_n(mx)]'}{j_n(mx)[xh_n^{(1)}(x)]' - h_n^{(1)}(x)[mxj_n(mx)]'};$$

$$c_n = \frac{j_n(x)[xh_n^{(1)}(x)]' - h_n^{(1)}(x)[xj_n(x)]'}{j_n(mx)[xh_n^{(1)}(x)]' - h_n^{(1)}(x)[mxj_n(mx)]'};$$

$$d_n = \frac{mj_n(x)[xh_n^{(1)}(x)]' - mh_n^{(1)}(x)[xj_n(x)]'}{m^2 j_n(mx)[xh_n^{(1)}(x)]' - h_n^{(1)}(x)[mxj_n(mx)]'}.$$

In fact, the former two series of scattering parameters, namely $a_n$ and $b_n$ are essential for the calculation of scattering and extinction cross-section.

In addition, $m$ is the refraction index of the sphere scatterer relative to the surrounding medium. Generally speaking, the refraction index of arbitrary medium is a complex number $n+ik$. The real part $n$ is equal to the ratio between vacuum light speed and the phase velocity of light in medium, and the imaginary part $k$ stands for the absorption of light. Furthermore, the refractive index of the same medium to different wavelengths of light is also different, which is called the dispersion of the medium. $n$ and $k$ are usually called the optical constants of matter.

2.1.2 Several important parameters

Define $\sigma_{\text{abs}}(\lambda, d)$, $\sigma_{\text{sca}}(\lambda, d)$, and $\sigma_{\text{ext}}(\lambda, d) = \sigma_{\text{abs}}(\lambda, d) + \sigma_{\text{sca}}(\lambda, d)$ to be the absorption, scattering and extinction cross-section, respectively, of a copper-containing particle with diameter d relative to the light of wavelength $\lambda$, and they are the functions of $\lambda$ and $d$. Theoretical scattering and extinction cross-section can be given based on the results above[20]:

$$C_{\text{sca}} = \frac{W_s}{I_i} = \frac{2\pi}{k^2} \sum_{n=1}^{\infty} (2n+1)\left(|a_n|^2 + |b_n|^2\right),$$

$$C_{\text{ext}} = \frac{W_{\text{ext}}}{I_i} = \frac{2\pi}{k^2} \sum_{n=1}^{\infty} (2n+1)\text{Re}(a_n + b_n).$$

and the difference between extinction and scattering cross-section is the absorption cross-section.

These cross-sections can be written as the product between corresponding dimensionless coefficients and the projection areas of these particles, namely $\sigma_{abs}(\lambda, d) = Q_{abs}(\lambda, d) \times \pi d^2/4$, $\sigma_{sca}(\lambda, d) = Q_{sca}(\lambda, d) \times \pi d^2/4$ and $Q_{ext}(\lambda, d) = Q_{abs}(\lambda, d) + Q_{sca}(\lambda, d)$. If we define $x=ka$, where $k$ is the wave vector, $a$ is the radius of spherical particle, then the corresponding dimensionless scattering and extinction coefficients can be defined as:

$$Q_{sca} = \frac{2}{x^2} \sum_{n=1}^{\infty} (2n+1)\left(|a_n|^2 + |b_n|^2\right),$$

$$Q_{ext} = \frac{2}{x^2} \sum_{n=1}^{\infty} (2n+1) \operatorname{Re}(a_n + b_n).$$

In this work, these three dimensionless coefficients can be calculated by MATLAB based on Mie scattering theory for any given wavelength and particle diameter.[20]

2.2 Basic assumptions

In order to highlight the main research object, and simplify the following calculations, some assumptions are made first:

(1) The reflection rate of the upper glaze surface is a constant $r_1$ ($0 \leq r_1 \leq 1$) for all the wavelengths;
(2) The reflection rate of the ceramic body is a constant $r_2$ ($0 \leq r_2 \leq 1$) for all the wavelengths;
(3) The copper-containing particles are uniformly dispersed in each transparent glaze layer paralleled to the glaze surface;
(4) The absorption character of pure transparent glaze satisfies the Lambert-Bill law. Assume the light intensity decreases to be $s_0 < 1$ after passing the whole transparent glaze layer. If the transparent glaze layer is uniformly divided into $N$ thin layers, then the light intensity will reduce to $s = \sqrt[N]{s_0}$ of that after passing each thin layer.
(5) The incident light is perpendicular to the glaze, and the scattering light is simplified as forward scattering and backward scattering terms;
(6) All the copper-containing particles are spheres.

In the following we will discuss the calculation details based on assumptions above.

2.3 Interaction between uniformly distributed copper-containing particles and light

Copper-containing particles will absorb and scatter the incident light of any wavelength $\lambda$. Generally, the light can be scattered to any direction, but for simplicity in our system the scattering light is simplified as forward scattering and backward scattering, and the forward/backward ratio is defined as the ratio of the power of forward and backward light scattered by a spherical particle.

Here the "forward" means that the angle between scattered light and incident light lies in 0~90°, and the "backward" means that the angle between scattered light and incident light lies in 90~180°. The forward/backward ratio $R_{fb}$ can be calculated by integrating the power of scattered light under different scattering angles.

Define the volume fraction of copper-containing particles in the glaze as $\omega$, then the particle number in per unit volume (namely the number density) is

$$\rho = 6\omega/\pi d^3$$

Under the assumption that the particles are uniformly dispersed, we can divide the glaze layer into $N$ thin layers along the depth direction, and define t as the thickness of each layer. If the intensity of incident light is $I_0$, the absorbed intensity $I_{abs}$, forward scattered intensity $I_{scaf}$, and backward scattered intensity $I_{scab}$ are:

$$\begin{cases} I_{abs} = I_0[1 - \exp(-n\sigma_{abs}t)] = I_0[1 - \exp(-tC_a(\lambda, d))] \\ I_{scaf} = I_{sca}\dfrac{r}{1+r} = \dfrac{rI_0}{1+r}[1 - \exp(-tC_s(\lambda, d))] \\ I_{scab} = I_{sca}\dfrac{1}{1+r} = \dfrac{I_0}{1+r}[1 - \exp(-tC_s(\lambda, d))] \end{cases}$$

where $C_a(\lambda, d)$ and $C_s(\lambda, d)$ are the product of corresponding absorption and scattering cross-section and the particle number density, they stand for the absorption and scattering intensity of the glaze with unit thickness.

Then, the light intensity $I_f$ along the original direction, and light intensity $I_b$ along the backward direction are

$$\begin{cases} I_f = (I_0 - I_{abs} - I_{scab} - I_{scaf})s + I_{scaf} = C_{ff}(\lambda, d, t)I_0 \\ I_b = I_{scab} = C_{bb}(\lambda, d, t)I_0 \end{cases}$$

where

$$\begin{cases} C_{\text{ff}}(\lambda, d, t) = s[\exp(-tC_a(\lambda, d)) + \exp(-tC_s(\lambda, d)) - 1] + \dfrac{r[1 - \exp(-tC_s(\lambda, d))]}{1 + r} \\ C_{\text{bb}}(\lambda, d, t) = \dfrac{[1 - \exp(-tC_s(\lambda, d))]}{r + 1} \end{cases}$$

Now we will analyze the mathematical relationship of the light intensities in different thin layers. As shown in Fig. 2(a), if we number the uppermost glaze thin layer as 1, and the glaze thin layer contact with the body as $N$, then we can formally number the air as layer 0, and the ceramic body as layer $N+1$.

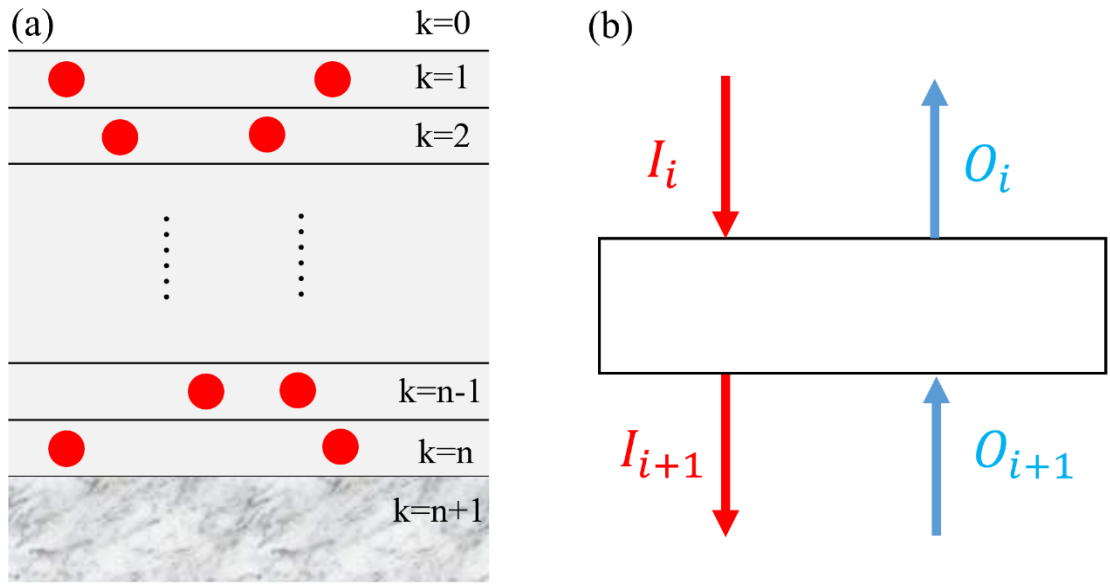

Fig. 2 Schematic diagram of the system with uniformly distributed copper-containing particles. (a) The transparent glaze is divided into $N$ thin layers, and the air and body can be formally regarded as layer 0 and layer $N+1$ respectively; (b) Incident and outgoing light from the upper and lower surfaces of the thin layer $i$.

For the thin layer numbered as $i$, if we define $I_i$ as the intensity of the incident light into the upper surface, $O_i$ as the intensity of the outgoing light away from the upper surface, as shown in Fig. 2(b), then the outgoing light $O_{i+1}$ away from the upper surface of thin layer $i+1$ will be the incident light into the lower surface of thin layer $i$, and the incident light $I_{i+1}$ into the upper surface of thin layer $i+1$ is also the outgoing light away from the lower surface of thin layer $i$.

Based on these relationships, the outgoing light intensity away from the thin layer $i$ can be written as:

$$O_i = C_{bb}(\lambda, d)I_i + C_{ff}(\lambda, d, t)O_{i+1}$$

Similarly, the outgoing light intensity away from the lower surface of thin layer $i$ can be written as:

$$I_{i+1} = C_{ff}(\lambda, d, t)I_i + C_{bb}(\lambda, d)O_{i+1}$$

The above two formulas are the relationships that the light should obey in the glaze, and the boundary conditions are also needed to determine the light intensity at any depth in the glaze.

First, at the body-glaze interface, according to assumption (2) we have $O_{N+1} = r_2 I_{N+1}$.

Second, assume the intensity of incident light into the upper surface to be 1, namely $I_0 = 1$. Taking the refraction and reflection at the glaze-air boundary into consideration, we have $O_0 = r_1 I_0 + (1 - r_1)O_1$, $I_1 = (1 - r_1)I_0 + r_1 O_1$.

Solving the question above is equivalent to solve a system of linear equations with 2N+4 components. Denote $I_k = x_{2k+1}$ and $O_k = x_{2k+2}$ (where k=0, 1, 2 ... N, N+1), then the six formulas above can be rewritten as:

$$\begin{cases} x_1 = 1 \\ r_1 x_1 - x_2 + (1 - r_1)x_4 = 0 \\ (1 - r_1)x_1 - x_3 + r_1 x_4 = 0 \\ C_{bb}(\lambda, d, t)x_{2k+1} - x_{2k+2} + C_{ff}(\lambda, d, t)x_{2k+4} = 0 \ (k > 0) \\ C_{ff}(\lambda, d, t)x_{2k+1} - x_{2k+3} + C_{bb}(\lambda, d, t)x_{2k+4} = 0 \ (k > 0) \\ r_2 x_{2N+3} - x_{2N+4} = 0 \end{cases}$$

The solved $x_2$ is the outgoing light intensity away from the glaze surface to air. The reflection spectrum of the transparent glaze model can be obtained after calculating a series of outgoing light intensities with different wavelengths.

2.4 Estimation of model parameters

According to Fresnel formula, the reflection rate of incident natural light at the glaze-air interface is

$$r_1 = \frac{1}{2}\left\{\left[-\frac{\sin(\theta_1 - \theta_2)}{\sin(\theta_1 + \theta_2)}\right]^2 + \left[\frac{\tan(\theta_1 - \theta_2)}{\tan(\theta_1 + \theta_2)}\right]^2\right\}$$

where $\theta_1$ and $\theta_2$ are incident angle and refraction angle respectively, and they satisfy refraction law $\sin\theta_1 = n_0 \sin\theta_2$, where $n_0$ is the refraction index, and can be reduced into $r_1 = \left(\frac{n_0-1}{n_0+1}\right)^2$ under the normal incidence condition. Ancient copper red glaze is a typical high temperature calcium alkali glaze, and its refraction is around 1.5. In this work we choose 1.52, and the corresponding reflection rate $r_1$ is about 0.04.

The whiteness of white porcelain body in Ding kiln can exceed 70%. As an estimation, the reflectivity $r_2$ of the body is taken as 0.7.

The absorption effect of transparent glaze is obviously related to the thickness and type of glaze. As an estimation, we assume that the light attenuates to 90% of the original intensity after passing through the transparent glaze without copper-containing particles, which is to say, $s_0$ is taken as 0.90.

In the computational simulation, the glaze thickness is set to 200 μm, which is the typical thickness of the red layer of copper red glaze observed in our previous experiment; the thickness $t$ of each thin layer is 5 times of the particle diameter, which can ensure that the interaction between particles is small, and then meet the far-field conditions of Mie scattering as much as possible.

As for the properties of colorant nanoparticles, previous researches have revealed that the nanoparticles in copper red glaze can be circular or polygonal in shape, and their diameters mainly fall in the interval of 50-100nm[1,3]. Therefore, in our simulation we choose the diameters to be 20, 50, 100 and 200nm to fit previous experimental results.

It should be pointed out here that the numerical calculation framework given above has many similarities with the classical four-flux model[21, 22] in terms of physical ideas, and the four-flux model can give analytical solutions. However, the advantage of our model is that it can be easily extended to non-uniform systems, such as the common multilayer glaze of ancient ceramics, and the system in which the particle size and number density of scattering particles change with depths in some middle-east areas[8,23].

2.5 CIE-L*a*b* color space

In order to describe the color of a particular object, the best way is measuring its reflection spectrum, but it is really inconvenient. Considering the fact that there are three kinds of optic cone cells, which are sensitive to blue, green and red colors respectively, the so-called tristimulus values X, Y and Z are introduced to describe the responses of these three kinds of cells to reflected light:

$$\begin{cases} X = k\sum_{\lambda} R(\lambda)S(\lambda)\bar{x}(\lambda)\Delta\lambda \\ Y = k\sum_{\lambda} R(\lambda)S(\lambda)\bar{y}(\lambda)\Delta\lambda \\ Z = k\sum_{\lambda} R(\lambda)S(\lambda)\bar{z}(\lambda)\Delta\lambda \end{cases}$$

where $R(\lambda)$ is the reflected intensity coefficient, $S(\lambda)$ is the relative spectral power distribution of standard light source, and $\bar{x}(\lambda)$, $\bar{y}(\lambda)$ and $\bar{z}(\lambda)$ are color matching functions of standard colorimetric observer, and a set of (X, Y, Z) values corresponds to a certain color.

Based on tristimulus values X, Y and Z, several different color space can be defined such as RGB, L*c*h*, and L*a*b*. The color space L*a*b* used in this manuscript is defined as:

$$\begin{cases} L^* = 116f\left(\dfrac{Y}{Y_n}\right) - 16 \\ a^* = 500\left(f\left(\dfrac{X}{X_n}\right) - f\left(\dfrac{Y}{Y_n}\right)\right) \\ b^* = 200\left(f\left(\dfrac{Y}{Y_n}\right) - f\left(\dfrac{Z}{Z_n}\right)\right) \end{cases}$$

where

$$f(t) = \begin{cases} \sqrt[3]{t} \\ \dfrac{t}{3\delta^2} + \dfrac{4}{29} \end{cases} \left(\text{if } t \le \delta^3,\ \delta = \dfrac{6}{29}\right)$$

and $X_n = 95.0489$, $Y_n = 100$, $Z_n = 108.8840$.

Contrary to (X, Y, Z) values, (L*, a*, b*) values are more intuitive: L* presents the brightness, and the object looks brighter when L* is larger; a* depict the red-green hue, the object looks redder when a* is larger, and greener when a* is smaller; b* depict the yellow-blue hue, the object looks yellower when b* is larger, and bluer when b* is smaller. L* ranges from 0 to 100, and both a* and b* range from –128 to +127.

## III. RESULTS AND DISCUSSION

3.1 Metallic copper particles as scatterers

Fig. 3 shows the calculation results of scattering coefficient $Q_{sca}$, forward/backward scattering rate $R_{fb}$ and absorption coefficient $Q_{abs}$ of metallic copper particles with different particle sizes for different wavelengths of light. The optical constant data used is taken from the paper of Johnson and Christy[24].

It is not difficult to see that for copper particles with diameters in the range of 20-200 nm, the peak position of scattering coefficient $Q_{sca}$ appears in the range of 600-700 nm, and the scattering coefficient of copper particles with diameter 100 nm for light with wavelength near 620 nm is close to 6. The variation trend of absorption coefficient $Q_{abs}$ of copper particles with different particle sizes for light with wavelength is roughly the same: for the light with wavelength below 600nm, the absorption coefficient is larger, while near 600nm, the absorption coefficient decreases precipitously.

The forward/backward scattering rate $R_{fb}$ of copper particles with different particle sizes is obviously different, but the variation trend is also obvious. Theoretically, when the particle size is much smaller than the wavelength of incident light, the forward and backward scattering intensities are equal, which corresponds to the Rayleigh scattering. Therefore, for particles with diameters of 20 nm and 50 nm, the forward/backward scattering rate $R_{fb}$ is basically equal to 1; for particles with larger size, the forward/backward scattering rate $R_{fb}$ first decreases rapidly with the increase of incident light wavelength, and then oscillates near $R_{fb}$ =1. Therefore, for some specific wavelengths and particle sizes, the light scattered backward will have higher intensity than the light scattered forward.

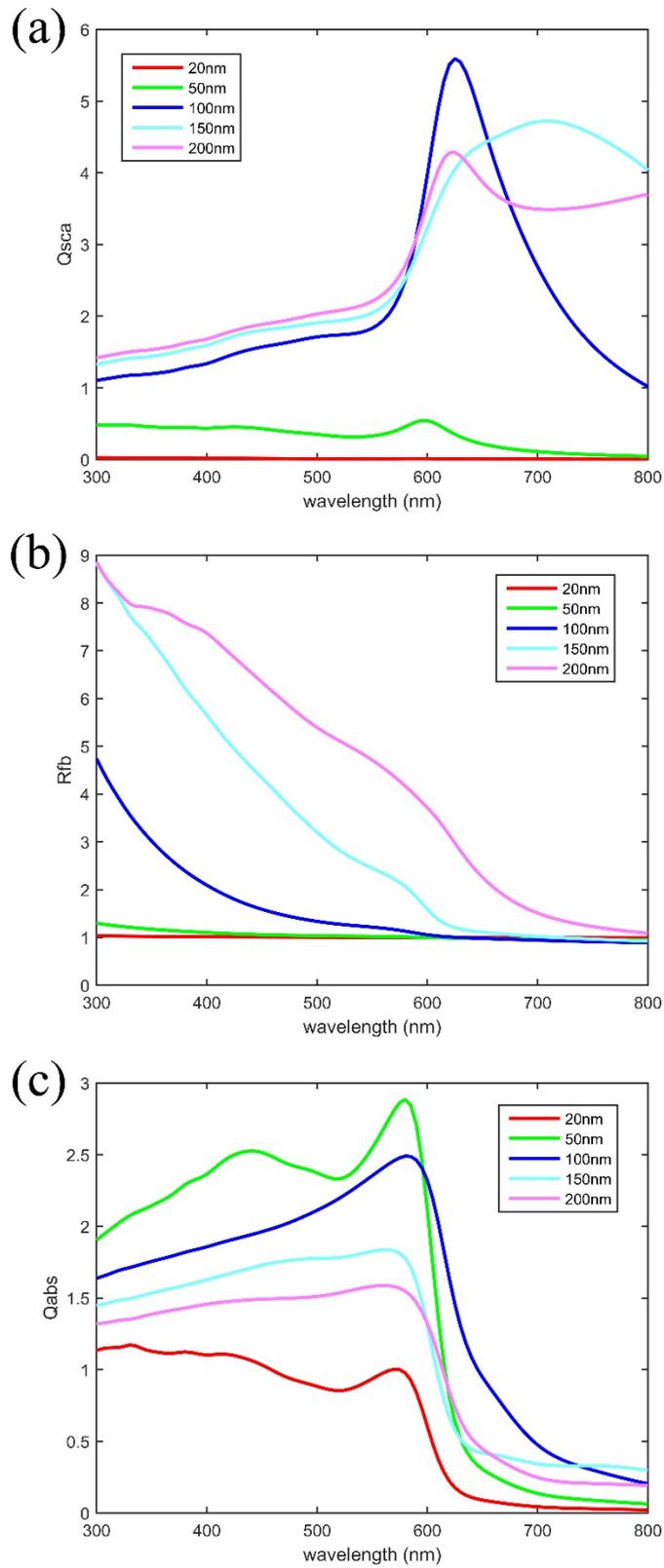

Fig. 3 (a) Scattering coefficient $Q_{sca}$ (b) forward/backward scattering rate $R_{fb}$ and (c) absorption coefficient $Q_{abs}$ of metallic copper particles with different particle sizes for light with different wavelengths.

Based on the parameters above, we calculated the reflection spectra of metallic copper particles with particle sizes of 50nm, 100nm and 200nm under different volume fractions, as shown in Fig. 4. Generally speaking, its common feature is that the reflectivity is low in the range of 300-600nm, while increasing rapidly and stabilizing at another higher value after passing through an interval of less than 50nm. Such a feature is also consistent with the measured reflection spectra of copper red glaze previously reported[13, 25].

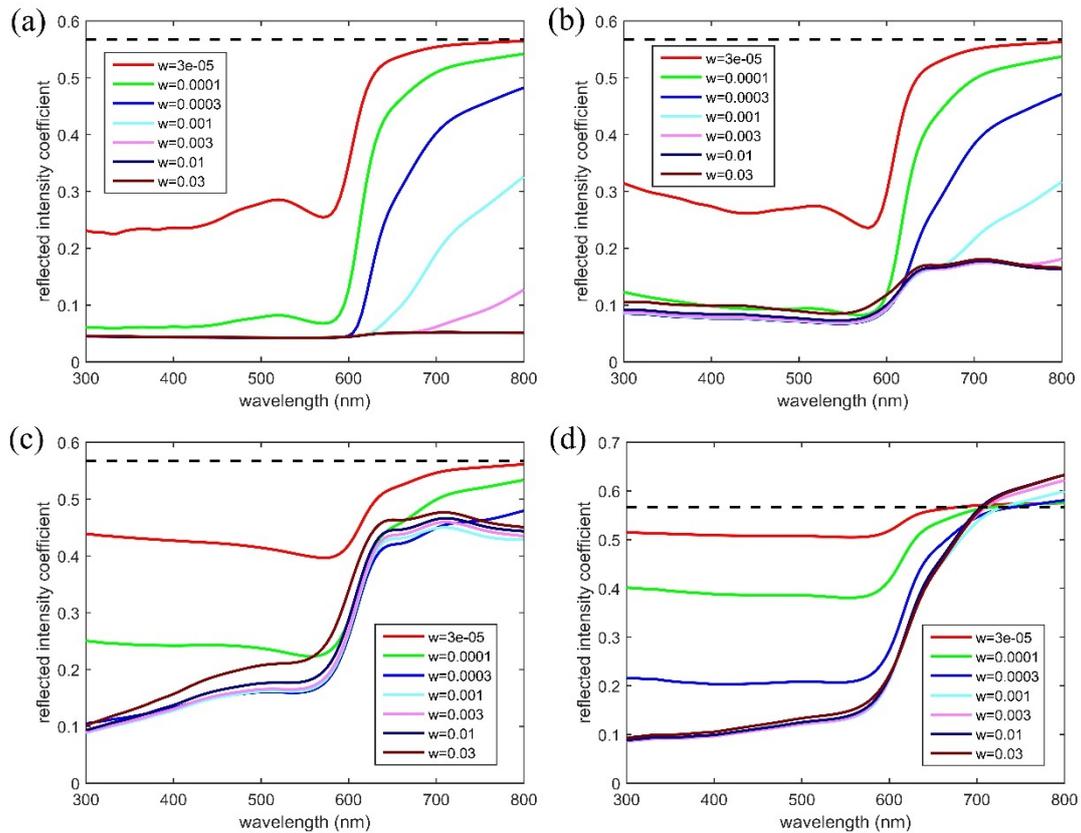

Fig. 4 Reflection spectra of metallic copper particles with diameters of (a) 20nm (b) 50nm (c) 100nm (d) 200nm under different volume fractions. The black dotted lines represent the reflection spectra of pure transparent glaze without any nanoparticles.

Furthermore, with the help of colorimetry, we can calculate the corresponding L*a*b* values according to the simulated reflection spectrum, which can objectively and quantitatively characterize the color of objects. Generally speaking, the larger the a* value is, the more obvious the red is. Table 1 shows the corresponding calculation results and recreated colors for each simulation condition. If a* > 15 is taken as the criterion for a good coloration effect, it can be seen from Table 1 and Fig. 4 that when the diameter of copper particles is 50nm, good coloration effect can be achieved when the volume fraction is in the range of 0.003%–0.03%, and the effect becomes worse when the content is higher; when the diameter is 100 nm, good coloration effect can be achieved when the volume fraction exceeds 0.03%; when the particle size is 200 nm, the volume fraction of particles needs to be at least 0.1%.

Table 1 L*a*b* values of reflection spectra produced by metallic copper particles

| Volume fraction | | 0.00003 | 0.0001 | 0.0003 | 0.001 | 0.003 | 0.01 | 0.03 |
|---|---|---|---|---|---|---|---|---|
| 20nm | L* | 61.6 | 39.2 | 28.2 | 25.0 | 24.6 | 24.6 | 24.7 |
| | a* | 11.1 | 27.0 | 19.7 | 3.9 | 1.5 | 1.5 | 1.5 |
| | b* | 7.5 | 13.8 | 6.2 | 0.8 | 0.1 | 0.1 | 0.05 |
| | color** | 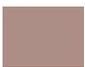 | 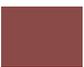 | 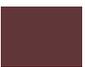 | 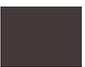 | 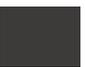 | 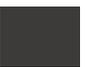 | 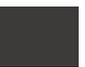 |
| 50nm | L* | 60.2 | 40.3 | 34.6 | 33.9 | 34.2 | 35.0 | 37.6 |
| | a* | 10.7 | 22.1 | 15.4 | 11.0 | 11.0 | 10.8 | 10.2 |
| | b* | 3.0 | 6.2 | 3.0 | 1.6 | 1.5 | 1.4 | 0.8 |
| | color | 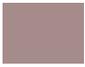 | 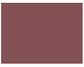 | 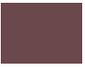 | 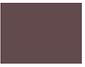 | 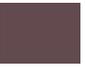 | 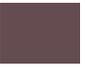 | 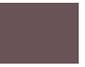 |
| 100nm | L* | 70.5 | 57.4 | 51.3 | 51.7 | 52.2 | 53.6 | 57.4 |
| | a* | 4.1 | 12.0 | 17.4 | 17.7 | 17.7 | 17.2 | 15.5 |
| | b* | -0.8 | 1.9 | 9.6 | 10.5 | 10.5 | 10.9 | 11.8 |
| | color | 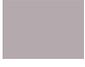 | 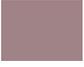 | 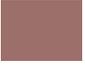 | 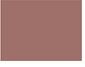 | 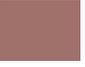 | 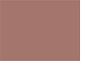 | 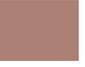 |
| 200nm | L* | 76.8 | 69.3 | 55.6 | 47.0 | 47.3 | 47.8 | 48.7 |
| | a* | 1.8 | 5.7 | 13.8 | 19.5 | 19.6 | 19.1 | 17.2 |
| | b* | 0.3 | 1.4 | 5.6 | 13.1 | 13.6 | 13.5 | 13.0 |
| | color | 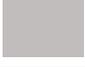 | 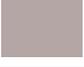 | 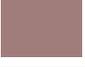 | 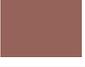 | 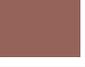 | 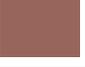 | 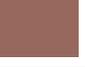 |

(**: the colors are recreated by https://www.qtccolor.com/secaiku/tool/convert?m=lab)

3.2 Cuprous oxide ($Cu_2O$)

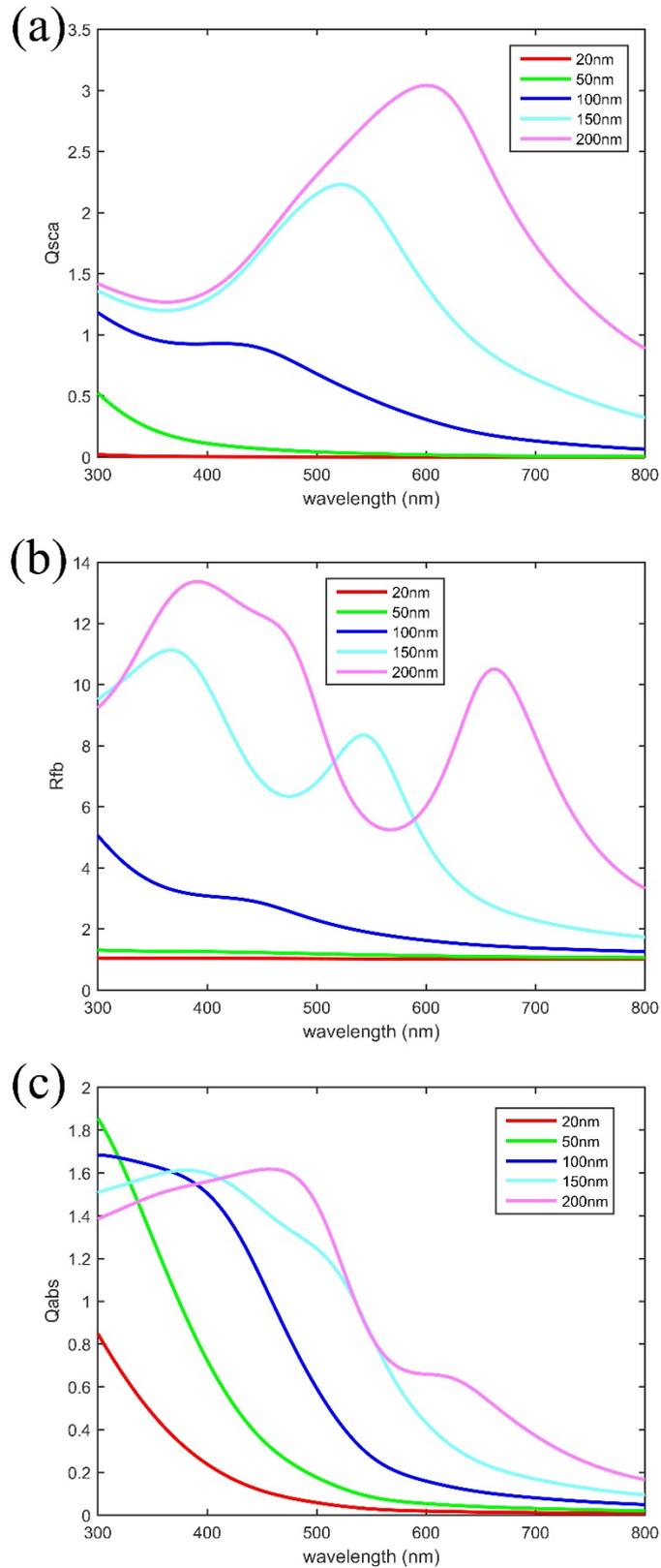

Fig. 5 (a) Scattering coefficient $Q_{sca}$ (b) forward/backward scattering rate $R_{fb}$ and (c) absorption coefficient $Q_{abs}$ of cuprous oxide particles with different diameters for light of different wavelengths.

Fig. 5 shows the calculation results of scattering coefficient $Q_{sca}$, forward/backward scattering rate $R_{fb}$ and absorption coefficient $Q_{abs}$ of cuprous oxide particles with different particle sizes for different wavelengths of light. The optical constants used in the calculation are taken from the monograph edited by Palik[26].

Different from metallic copper particles, for $Cu_2O$ particles the peak position of scattering coefficient $Q_{sca}$ moves from about 300 nm to about 600 nm with the increase of particle size from 50 nm to 200 nm, and the maximum value of scattering coefficient is about 3. The absorption coefficient $Q_{abs}$ of $Cu_2O$ particles with different diameters basically decreases with the increase of wavelength, and the change is relatively smooth, without precipitously decrease as shown by copper particles at 600nm. In addition, the change of forward/backward scattering rate $R_{fb}$ of $Cu_2O$ particles with different diameters is more complex. For example, for $Cu_2O$ particles with diameters of 150 nm and 200 nm, there are two maxima and one minimum in the range of 300-800 nm.

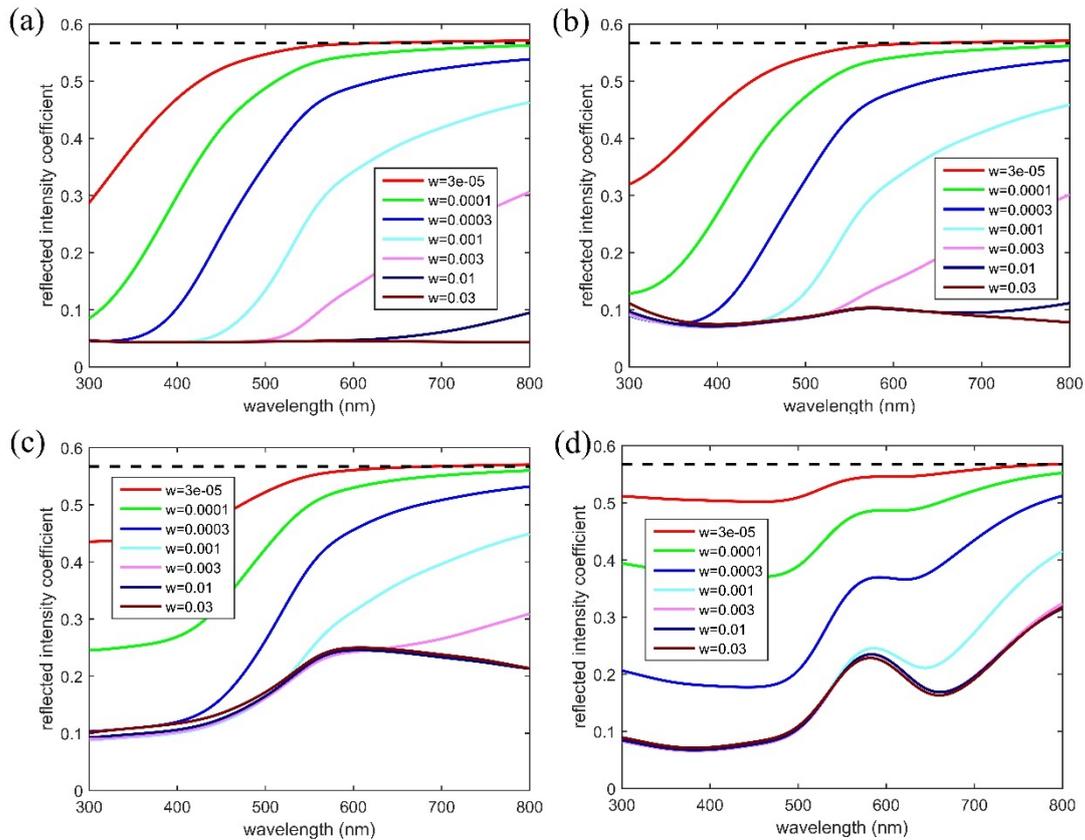

Fig. 6 Reflection spectra of cuprous oxide particles with diameters of (a) 20nm (b) 50nm (c) 100nm (d) 200nm under different volume fractions. The black dotted lines represent the reflection spectra of pure transparent glaze without any nanoparticles.

On this basis, we further calculated the reflection spectra of cuprous oxide particles with diameters of 50nm, 100nm and 200nm under different volume fractions. As shown in Fig. 6, there is a significant difference between the reflection spectra generated by cuprous oxide and metallic copper particles. Although the reflection spectra also increase with the increase of wavelength, the intervals of the rising section are obviously wider than that of the corresponding reflection spectra of metallic copper particles, and their position gradually redshifts with the increase of particle size.

Table 2 L*a*b* values of reflection spectra produced by cuprous oxide particles

| Volume fraction | | 0.00003 | 0.0001 | 0.0003 | 0.001 | 0.003 | 0.01 | 0.03 |
|---|---|---|---|---|---|---|---|---|
| 20nm | L* | 79.5 | 77.3 | 71.5 | 57.0 | 36.2 | 25.4 | 25.2 |
| | a* | -0.5 | -1.1 | -0.3 | 9.4 | 15.9 | 1.0 | 0 |
| | b* | 3.7 | 11.4 | 27.7 | 44.3 | 19.6 | 1.1 | 0.7 |
| | color** | 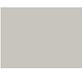 | 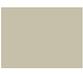 | 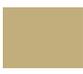 | 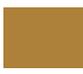 | 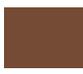 | 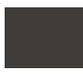 | 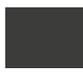 |
| 50nm | L* | 79.3 | 76.8 | 70.5 | 56.0 | 40.8 | 37.0 | 37.2 |
| | a* | -0.6 | -1.2 | 0.5 | 10.9 | 9.2 | -0.2 | -0.2 |
| | b* | 4.5 | 13.7 | 30.7 | 36.0 | 13.0 | 6.1 | 5.8 |
| | color |  | 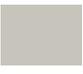 | 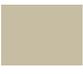 | 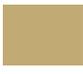 | 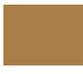 | 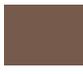 | 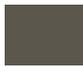 |
| 100nm | L* | 78.7 | 75.1 | 67.4 | 56.3 | 52.7 | 53.1 | 53.7 |
| | a* | -0.3 | 0.1 | 4.5 | 9.0 | 2.7 | 2.1 | 2.1 |
| | b* | 6.3 | 18.1 | 32.9 | 25.1 | 18.4 | 18.1 | 16.7 |
| | color | 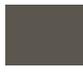 |  | 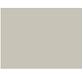 | 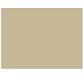 | 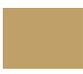 | 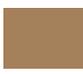 | 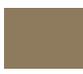 |
| 200nm | L* | 78.0 | 72.9 | 62.2 | 50.2 | 49.2 | 49.3 | 49.1 |
| | a* | 0.9 | 2.7 | 6.0 | 5.5 | 3.6 | 3.4 | 2.4 |
| | b* | 3.0 | 9.3 | 21.6 | 28.2 | 27.0 | 26.6 | 24.8 |
| | color | 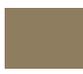 | 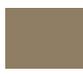 |  | 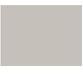 | 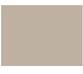 | 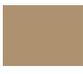 | 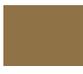 |

(**: the colors are recreate by https://www.qtccolor.com/secaiku/tool/convert?m=lab)

Table 2 shows the L*a*b* values and recreated colors of the system under different conditions when cuprous oxide particles are used as scatterers, as well as recreated colors for each simulation condition. It is not difficult to find that except the case that $d$=20nm, $\omega$=0.3%, the a* value of the corresponding reflection spectra in other cases is less than 15; that is to say, it is not red enough. What is more, the relationship b*> a* is common in Table 2, which means that when the scatterer is cuprous oxide, the system will intuitively appear as a mixed color of red and yellow rather than bright red.

3.3 Comparison with experimental results

As a comparison, we found the measured chromaticity value of sacrificial red glaze in the Ming and Qing Dynasties in the literatures, as shown in Table 3 below.

Table 3 measured chromaticity value of sacrificial red glaze in Ming and Qing Dynasties

| Dynasty | Sample number | L | a | b |
| --- | --- | --- | --- | --- |
| Ming[27] | YL-1-1 | 28.4 | 13.9 | 6.4 |
| | YL-1-2 | 26.5 | 20.1 | 8.2 |
| | YL-1-3 | 20.4 | 23.1 | 7.7 |
| | XD-1-1 | 26.2 | 20.2 | 7.7 |
| | XD-1-2 | 12.3 | 17.4 | 5.7 |
| | XD-1-3 | 22.5 | 24.9 | 7.8 |
| Qing[27, 28] | KX-1 | 28.5 | 23.9 | 7.8 |
| | YZ-1 | 19.3 | 16.0 | 4.8 |
| | QL-1 | 27.5 | 16.4 | 5.7 |
| | QL-2* | 14.7 | 11.7 | 4.0 |
| | DG-1* | 22.8 | 12.9 | 4.5 |

(*: the colors of these two samples are pig liver-colored and dark red, which are not representative enough)

In order to facilitate understanding, the data in the above four tables can be drawn in the same figure. Considering that the values of a* and b* are more closely related to coloration, we choose b* as the horizontal coordinate and a* as the vertical coordinate to draw the scatter diagram, as shown in Fig. 7.

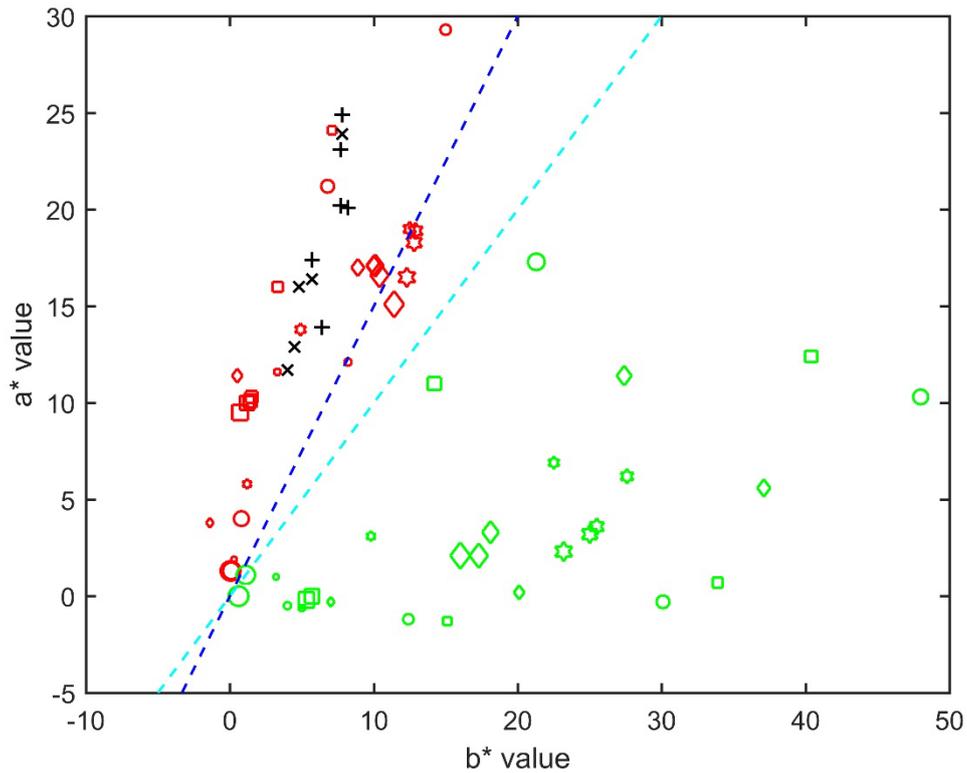

Fig. 7 Chromaticity a* and b* values corresponding to simulated reflection spectra and experimental results. The red and green colors represent the case where the scatterers are metallic copper and cuprous oxide particles respectively; the round, square, diamond and hexagonal shapes represent the diameters of 20nm, 50nm, 100nm and 200nm respectively, and the figures from small to large represent seven different volume fractions. In addition, the "+" in black represents the measured chromaticity value of sacrificial red glaze of the Ming Dynasty, while the "x" in black represents the measured chromaticity value of the sacrificial red glaze of the Qing Dynasty. Blue dotted lines is a*=3b*/2, and cyan dotted line is a*=b*.

It is not difficult to see from Fig. 7 that when the scatterers are metallic copper particles, their a* values basically fall on the left side of the blue dotted line a*=3b*/2, and all of them fall on the left side of cyan dotted line a*=b*, which is in sharp contrast to $Cu_2O$ particles. Moreover, the measured chromaticity values of sacrificial red glaze samples also fall in the corresponding area of metallic copper particles, which further proves that sacrificial red glaze mainly depends on metallic copper particles as colorant.

It should be noted that whether there are $Cu_2O$ particles in the glaze cannot be determined just based on the results of reflection spectrum. In previous literature, researchers judged the coloration mechanism simply by analyzing the phase of crystalline particles in glaze, which may not be reliable. According to the results of this paper, it is difficult for cuprous oxide particles as scatterers in transparent glaze to give bright red. Therefore, if the reflection spectrum of the sample shows the characteristics of typical copper red glaze, that is to say, it rises rapidly near 600nm and the chromaticity value $a^*>3b^*/2$, then it can be considered that its colorant is mainly the metallic copper particles, even if both metallic copper particles and cuprous oxide particles do exist in the system simultaneously. If the increasing interval of the reflection spectrum is wide and the chromaticity value $b^*>a^*$, cuprous oxide is likely to play a role in coloration.

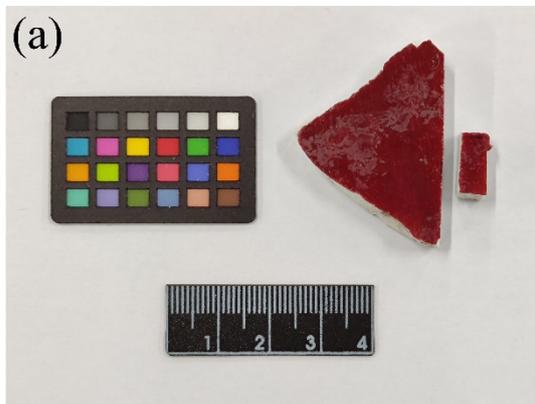
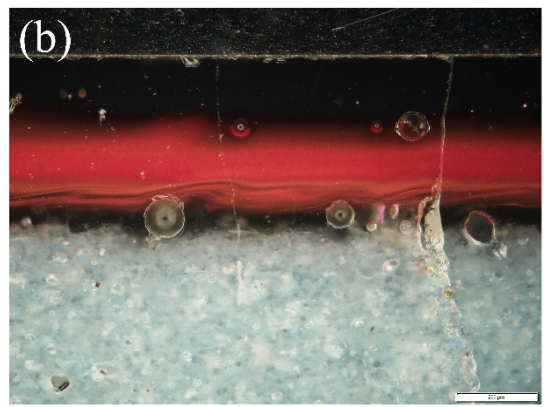
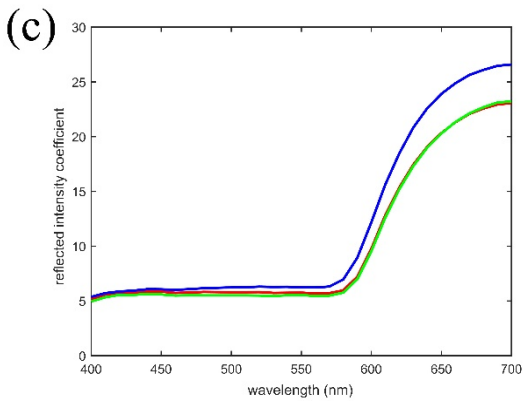
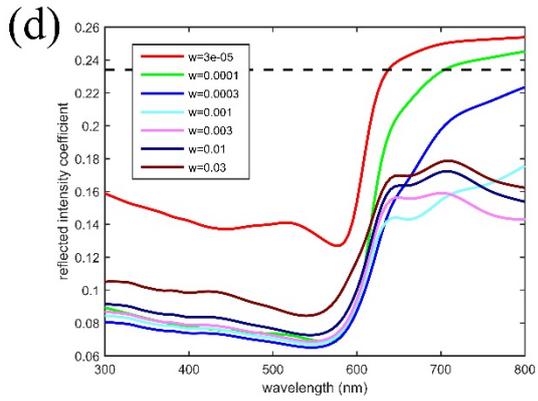
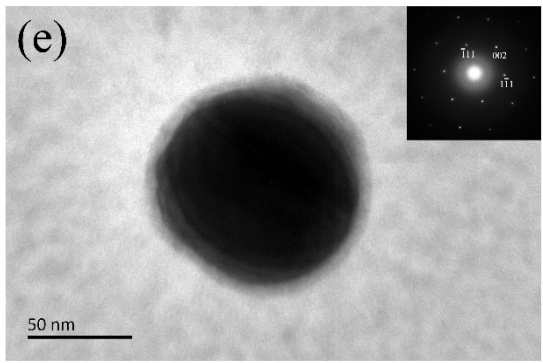
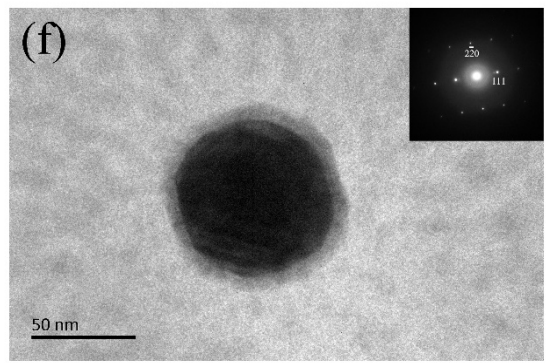

Fig 8 (a) furnace transmutation copper red glaze sample unearthed from Nandaku site in the Palace Museum; (b) dark field image of the cross-section sample; (c) three experimental reflection spectra; (d) simulated reflection spectra of different volume fractions, here the parameter $r_2$=0.65 and $s_0$=0.8; (e) and (f) are TEM images of copper nanoparticles and corresponding diffraction patterns.

As an application, we analyzed one furnace transmutation copper red glaze sample shown in Fig. 8(a), which was produced during the Qianlong reign period in the Qing Dynasty, and unearthed from the Nandaku Site in the Palace Museum. Fig. 8(b) shows the optical dark field image of the cross-section sample. It can be seen that the glaze can be divided into three layers: the top transparent layer, the red layer and the bottom transparent layer.

Fig. 8(c) shows three reflection spectra measured at three different points of the sample, which rise rapidly near 600nm. The averaged chromaticity values calculated are L*=33.0, a*=17.7 and b*=7.1, which also satisfies the relation a*>3b*/2. According to the theory just discussed above, the colorant should mainly be the metallic copper particles.

In fact, the reflection spectrum can also be simulated, as shown in Fig. 8(d), where the diameter of copper nanoparticles in simulation is 50nm. However, the other parameters used here are a little different from previous simulations: the measured reflectivity of this ceramic body is about 65%, so the reflection rate $r_2$ is chose as 0.65; meanwhile, the absorption coefficient $s_0$ is decreased from 0.90 to 0.6, because the total thickness of glaze is much larger than that of the red layer. It can be seen that the simulated reflection spectrum with $\omega$=0.0001 (green line) is quite similar to the experimental results, which indirectly prove that the colorant of this sample may be metallic copper nanoparticles with diameter around 50nm.

Direct evidence is given by TEM, as shown in Fig. 8(e) and Fig. 8(f), where the shapes of such nanoparticles are close to sphere with diameters of 50-100nm, and the diffraction patterns in the upper-right corner demonstrate that these nanoparticles are really pure metallic copper particles.

The reflection spectra were measured by colorimeter CHN Spec CS-520. The FIB cutting of the TEM sample was conducted with a Zeiss Auriga Compact dual beam instrument equipped with an Omniprobe AutoProbe 200 micromanipulator at IGGCAS (Institute of Geology and Geophysics, Chinese Academy of Science). Ion beam condition for the final thinning and polishing were 5–30kV high voltage with beam currents of 50pA–2nA. The FIB section was prepared to about 100nm and $10 \times 3\mu m^2$ in area. The TEM imaging and selected area electron diffraction were carried out using a JEOL JEM-2100 TEM operated at 200kV with Oxford X-MAX energy dispersive X-ray spectrometers at IGGCAS.

## IV. CONCLUSION

In this paper, using the computational simulation method based on multiple scattering, and aiming at the situation that the copper-containing particles are uniformly distributed in the copper red glaze, the reflection spectra of the glaze with different particle sizes, different volume fractions and different kinds of nanoparticles are calculated. The calculation results show that the metallic copper particles in the range of diameter from 20 nm to 200 nm may make the glaze appear as a good copper red glaze, but the volume fraction requires changes as the size changes. When cuprous oxide nanoparticles are used as scatterers, their spectral characteristics are obviously different from that produced by metallic copper particles, so it is difficult to achieve the effect of copper red glaze. The chromaticity value of sacrificial red glaze reported in the literature is consistent with our calculation results for metallic copper particles, and our experiments of copper red glaze sample also support our simulation results. In this sense, we can quickly judge the coloration mechanism by analyzing its reflection spectrum.

## ACKNOWLEDGEMENTS

**The authors would like to thank Jiayu Hou, An Gu, Yao Chen, An Gu and Dr. Ming Guan for their help for this work. We thank Xu Tang and Lixin Gu at the Electron Microscopy Laboratory, Institute of Geology and Geophysics, Chinese**

Academy of Science, for their efforts to maintain smooth operation in FIB-SEM & TEM experiments. This work was financially supported by the Chinese National Natural Science Foundation (No. U1832164, U1932203).

**REFERENCES**

[1] C. Jia, G. Li, M. Guan, J. Zhao, Y. Zheng, G. Y. Wang, X. J. Wei and Y. Lei, J. Eur. Ceram. Soc., **41**(6) 3809-3815 (2021).

[2] S. B. Tian, Y. Z. Liu, M. L. Zhang, L. H. Wang, Y. N. Xie and C. S. Wang, Nuclear Techniques, **32**(06), 413-418 (2009).

[3] J. Y. Hou, C. Jia, X. G. Zhang, H. Li, G. Li, H. W. Liu, B. Q. Kang and Y. Lei, J. Eur. Ceram. Soc., **42**(3), 1141-1148 (2022).

[4] Y. M. Luo, W. Pan, S. Q. Li, H. L. Zhao and J. Wang, Ceramics, (06), 23-26, (2000). (DOI: j.cnki.ceramics.2000.06.005)

[5] M. Wakamatsu, N. Takeuchi, H. Nagai, Y. Ono and S. Ishida, J. Ceram. Soc. Jpn., **94**(4), 387-392 (1986).

[6] J. Y. Hou, T. Pradell, Y. Li and J. M. Miao, J. Eur. Ceram. Soc., **38**(12), 4290-4302 (2018).

[7] T. Pradell, G. Molina, J. Molera, J. Pla and A. Labrador, Appl Phys A **111**, 121-127 (2013).

[8] G. Molina, M. S. Tite, J. Molera, A. C.- Font and T. Pradell, J. Eur. Ceram. Soc, **34**(10), 2563-2574 (2014).

[9] Y. Wang, S. H. Yu, M. H. Tong, W. W. Wang and X. Y. Yang, J. Cult. Herit., **48**, 29-35 (2021).

[10] J. W. Xue, J. W. Zhong, Y. R. Mao, C. H. Xu, W. Liu and Y. Q. Huang, Ceram. Int., **46**(14), 23186-23193 (2020).

[11] M.-L. Zhang, L.-H. Wang, L.-L. Zhang and H.-S. Yu, Nucl. Sci. Tech., **30**, 114 (2019).

[12] G. Chen, S. Q. Chu, T. X. Sun, X. P. Sun, L. R. Zheng, P. F. An, J. Zhu, S. R. Wu, Y. H. Du and J. Zhang, J. Synchrotron Rad., 24 (2017).

[13] C. Lu, J. X. Jiang, C. Q. Xu, S. R. Wu, L. Guan, J. Zhang, D. L. Chen, W. Xu, J. Zhuand C. S. Wang, Journal of University of Chinese Academy of Science, **33**(03),

421-426 (2016).

[14]Y. Q. Li, Y. M. Yang, J. Zhu, X. G. Zhang, S. Jiang, Z. X. Zhang, Z. Q. Yao and G. Solbrekken, Ceram. Int., **42**(7), 8495-8500 (2016).

[15]J. Zhu, H. P. Duan, Y. M. Yang, L. Guan, W. Xu, D. L. Chen, J. Zhang, L. H. Wang, Y. Y. Huang and C. S. Wang, J Synchrotron Rad., **21**, 751-755 (2014).

[16]I. Nakai, C. Numako, H. Hosono and K. Yamasaki, J Am. Ceram. Soc., **82**(3), 689-695 (1999).

[17]T. Pradell, R. S. Pavlov, P. C. Gutiérrez, A. C.-Font and J. Molera,. J. Appl. Phys., **112**, 54307 (2012).

[18]P. A. Cuvelier, C. Andraud, D. Chaudanson, J. Lafait and S. Nitsche, Appl Phys A, **106**, 915-929 (2011).

[19]C. F. Bohren and D. R. Huffman, Absorption and Scattering of Light by Small Particles. John Wiley & Sons, Inc. (1983).

[20]C. Mätzler, MATLAB Functions for Mie Scattering and Absorption (2002).

[21]B. Maheu, J. N. Letoulouzan and G. Gouesbet, Appl. Opt., **23**(19), 3353-3362 (1984).

[22]B. Maheu and G. Gouesbet, Appl. Opt., **25**(7), 1122-1128 (1986).

[23]P. Sciau, L. Noé and P. Colomban, Ceram. Int., **42**(14), 15349-15357 (2016).

[24]P. B. Johnson and R. W. Christy, Phys. Rev. B, **6**(12), 4370-4379 (1972).

[25]C. E. Cao, H. R. Shen, J. W. Cao, C. H. Xiong and N. Z. Zheng, China Ceramic Industry, **9**(6), 4-8 (2002).

[26]E. D. Palik, Handbook of Optical Constants of Solids II. Academic Press (1998).

[27]Q. Yu, Technology Study of the Qing Dynasty Sacrificial Red Glaze (2013).

[28]J. Wu, Q. Yu, M. L. Zhang, J. M. Wu, Q. J. Li and Y. F. Wu, Journal of Ceramics, **34**(04), 476-481 (2013).